\documentclass{svproc}
\usepackage{url}

\usepackage{graphicx}
\usepackage{multicol}
\usepackage{footmisc}

\begin{document}
\mainmatter              
\title{Lattice QCD Impact on Determination of the CKM Matrix}  
\titlerunning{Lattice QCD impact}  
%
\author{Steven Gottlieb}
\authorrunning{Steven Gottlieb} 
%
%
\institute{Indiana University, Bloomington IN 47405, USA,\\
\email{sg@indiana.edu},\\ WWW home page:
\texttt{http://physics.indiana.edu/\~{}sg/}
}

\maketitle              

\begin{abstract}
We review many lattice QCD calculations that impact the precise 
determination of the CKM matrix.
We focus on decay constants and semileptonic
form factors of both light ($\pi$ and K)
and heavy-light ($D_{(s)}$ and $B_{(s)}$) mesons.  Implication of
$\Lambda_b$ form factors will be shown.
When combined with experimental results for branching fractions and
differential decay rates, the above calculations strongly constrain
the first two rows of the CKM matrix.  We discuss a long standing
difference between $|V_{ub}|$ and $|V_{cb}|$ as determined from
exclusive or inclusive decays.
\end{abstract}
\section{Introduction}
Lattice QCD contributes strongly to understanding
the CKM matrix and the search for beyond the Standard Model (SM)
physics.  To begin, I would
like to relate a little about my background.  I am a member of the
Flavour Lattice Averaging Group (FLAG) and participate in the
$D$ and $B$ semileptonic decays working group.  However, this is not a
FLAG approved talk.  The most recent FLAG review~\cite{Aoki:2016frl} 
dates from 2016, and I will concentrate on more recent plots.
The closing date for papers to appear in the next review is a couple of months
after FPCP 2018, so the next FLAG plots and averages are not yet
available.

For more than a decade, my own research has been in the context
of the MILC collaboration and the Fermilab Lattice/MILC Collaborations 
(FNAL/MILC).
Much of our work is directed toward more precisely determining the 
CKM matrix, and looking for discrepancies that would indicate physics
beyond the SM, I will liberally use plots from my own
collaborations when they contain results not yet reviewed by FLAG.

I also happen to be a member of the local organizing committee for 
Lattice 2018, which takes place the week after FPCP 2018.
I can assure you that topics discussed here
are very active areas within the lattice QCD community.  For Lattice
2018, there were 37 abstracts in the weak matrix element category.  Of them,
five dealt with decay constants, 12 with $K$, $D$, or $B$ meson semileptonic
decay, and seven with nucleon or nuclear matrix elements.  
%
\subsection{CKM matrix}
In expression~\ref{eq:ckmmatrix},
CKM matrix elements are in bold and below each of them are one or two
processes that can be used to determine that element.
Below the last row, the matrix elements represent the
$B_{(s)} \bar B_{(s)}$ mixing phenomena that 
depend on $V_{td}$ and $V_{ts}$ through loop diagrams.

        \begin{eqnarray}
        \left(
        \begin{array}{ccc}
        { \bf{V_{ud}}}   &  \bf{ V_{us}}  &   \bf{ V_{ub} }\\
        \pi\to l\nu & K\to l\nu  & B\to l\nu \\
                    & K\to\pi l\nu  & B\to\pi l\nu \\
        \bf{ V_{cd} }  &  \bf{ V_{cs}  } &   \bf{ V_{cb}} \\
        D\to \pi l\nu & D\to K l\nu & B\!\to\! D^{(\!*\!)}\! l \nu \\
        D\to l\nu & D_s\to l\nu & \Lambda_b\to\Lambda_c l\nu  \\
        \bf{ V_{td}}  &\bf{ V_{ts}}  & \bf{ V_{tb}} \\
        \langle B_d | \overline{B}_d\rangle &
        \langle B_s | \overline{B}_s\rangle \\
        \end{array}
        \right)
\label{eq:ckmmatrix}
        \end{eqnarray}

The CKM matrix is unitary, so each row and column is a 
complex unit vector and each row (or column) is orthogonal to the 
other two.  Violation of unitarity would be evidence for non-SM
physics.  Since different decays can depend on the same CKM matrix
element, if the value of the matrix element inferred from the different
decays do not agree, that would be evidence for new physics.

Consider the branching fraction $\mathcal{B}$ for leptonic 
decay of a $D$ or $D_s$ meson.

\begin{equation}
{\mathcal{B}}(D_{(s)} \to \ell\nu_\ell)= {{G_F^2|V_{cq}|^2 \tau_{D_{(s)}}}\over{8 \pi}} f_{D_{(s)}}^2 m_\ell^2 
m_{D_{(s)}} \left(1-{{m_\ell^2}\over{m_{D_{(s)}}^2}}\right)^2
\end{equation}
where
$V_{cq}$ is the (unknown) CKM matrix element with $q=d$, or $s$, 
$f_{D_{(s)}}$ is the decay constant of the meson that we calculate using
lattice QCD, and the other factors on the RHS are well known.
For semileptonic decays, the LHS would be a differential decay rate, and
the RHS would involve a CKM matrix element and form factors describing the
transition matrix element between the initial and final state hadrons induced
by the weak current responsible for the decay.  In both cases, the experimental
measurement and hadronic input from lattice QCD allow determination of the CKM
matrix element.

\section{First Row: Light Quarks}
\begin{figure}[tb]
\begin{center}
\includegraphics[width=0.7\textwidth]{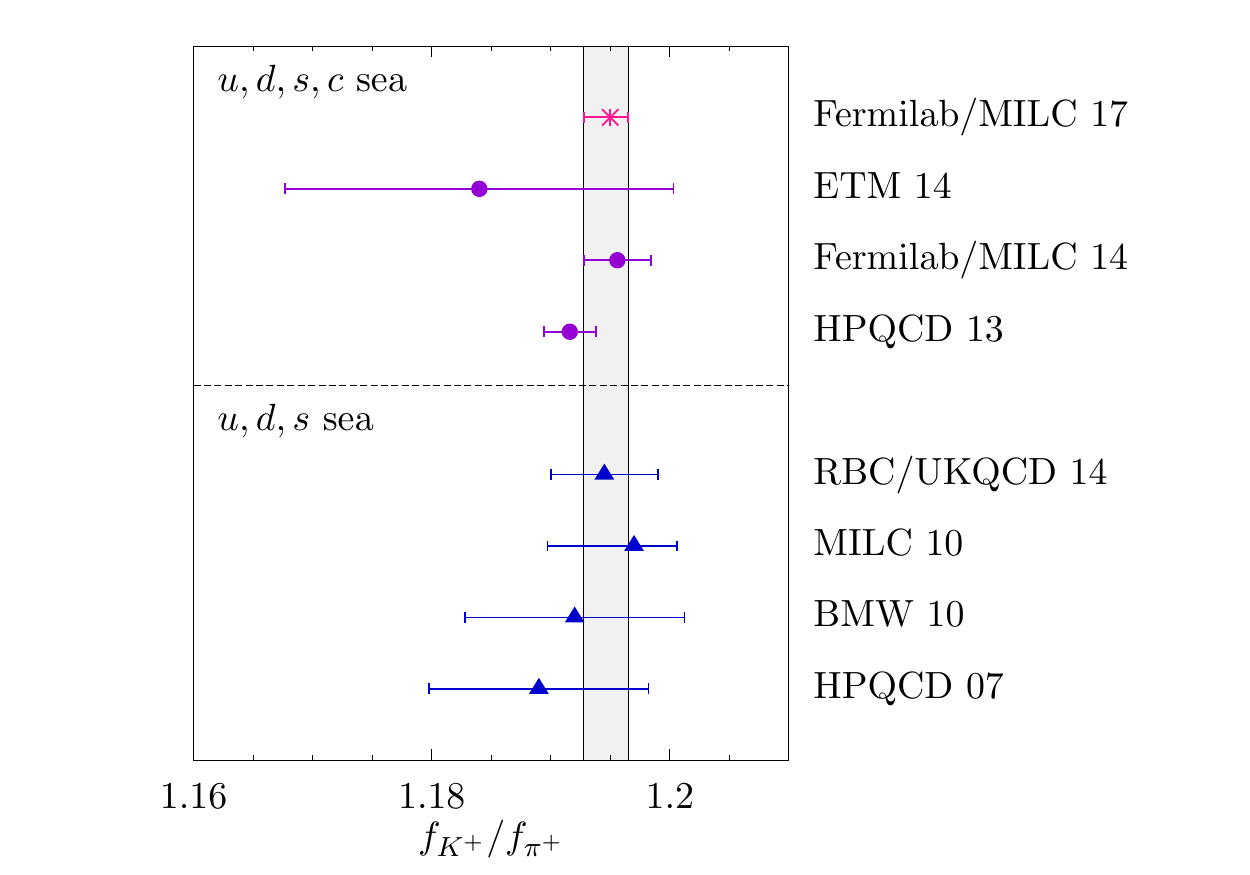}
\vspace{-5mm}
\caption{
\label{fig:fKoverFpi}
Comparison of calculations of decay constant ratio $f_K/f_\pi$ with
$N_f=2+1+1$ and $2+1$ sea quark flavors.  From \cite{Bazavov:2017lyh}.
}
\end{center}
\end{figure}

We start with decays of $\pi$ and $K$ mesons.  As implied
by \ref{eq:ckmmatrix} and subsequent discussion generalized to other
mesons, we need $f_\pi$ and $f_K$ to describe
the leptonic decays and a form factor to describe the kaon semileptonic
decay.  The two decay constants can each be calculated; however, $f_\pi$
is often used to set the lattice scale, so the ratio $f_{K^\pm}/f_{\pi^\pm}$
which has the advantage of smaller systematic errors is a key quantity.
From experiment~\cite{Patrignani:2016xqp} it is known that
\begin{equation}
\left|{V_{us}\over V_{ud}}\right| {f_{K^\pm}\over f_{\pi^\pm}} = 0.2760(4)\;.
\end{equation}
Thus, knowledge of the decay constant ratio, allows us to determine the
ratio of the first two elements of the CKM matrix.  The decay constant
ratio has recently been updated by FNAL/MILC~\cite{Bazavov:2017lyh}.  The
result is $f_{K^\pm}/f_{\pi^\pm} = 1.1950(^{+15}_{-22})$ which may
be compared with the FLAG 2016~\cite{Aoki:2016frl} $N_f=2+1+1$ average 1.193(3).
$N_f$ is the number of dynamical sea quarks in the calculation.
Figure~\ref{fig:fKoverFpi} summarizes the most relevant calculations 
including those with $N_f=2+1$.

\begin{figure}[thb]
\begin{center}
\includegraphics[width=0.60\textwidth]{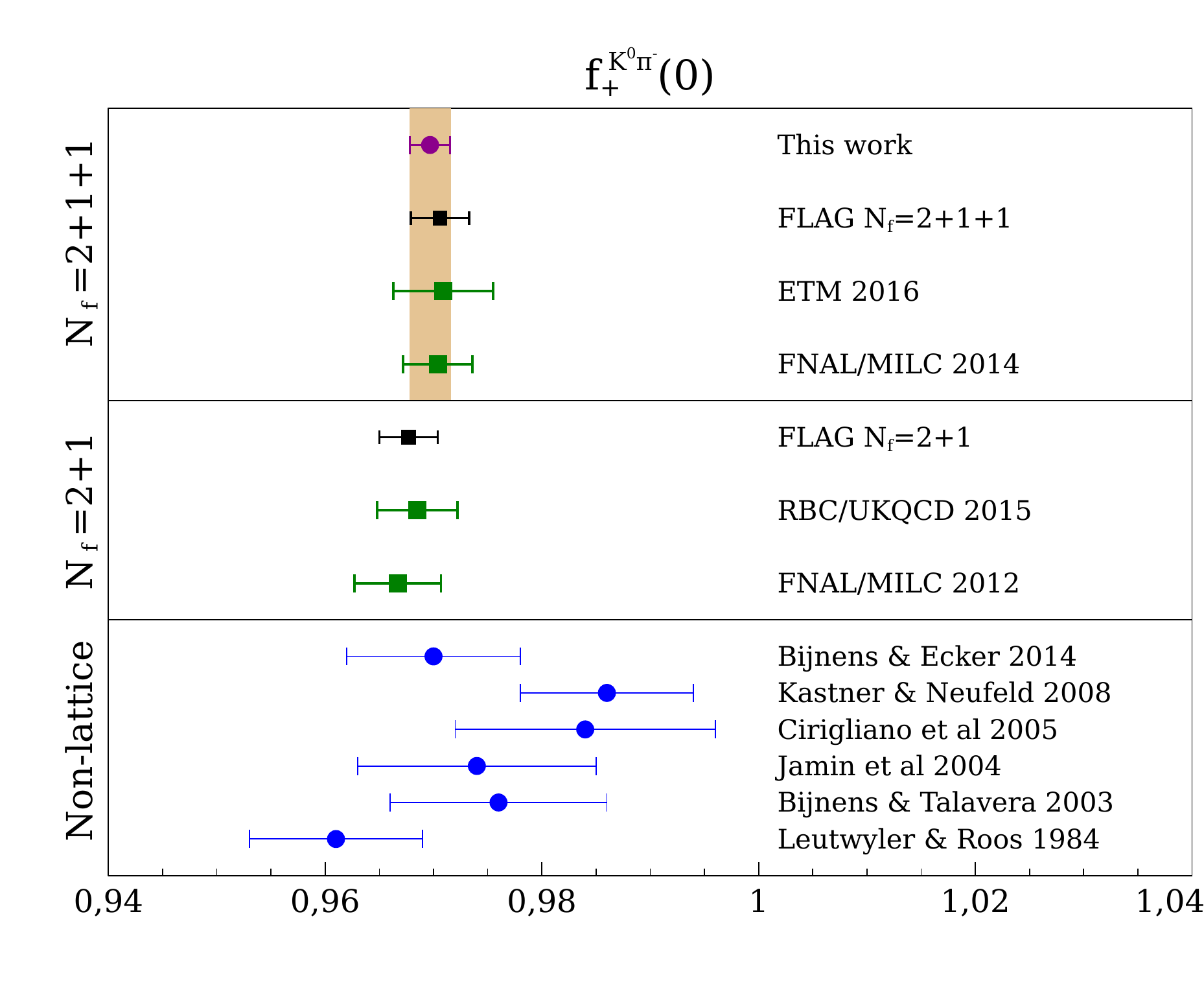}
\vspace{-5mm}
\caption{
\label{fig:Form_factors}
Comparison of calculations of the kaon form factor $f^K_+(q^2=0)$ with
$N_f=2+1+1$, or $2+1$ sea quark flavors, and non-lattice 
QCD calculations~\cite{Bazavov:2018kjg}.
}
\end{center}
\end{figure}

For kaon semileptonic decay, $p_K = p_\pi + q_\ell + q_\nu$ by energy-momentum
conservation.  The relevant variable for the form factors is $q^2$ with
$q = q_\ell + q_\nu$ the 4-momentum of the leptons.  One could,
in principle, determine the shape of the vector form factor $f_+(q^2)$
to predict the shape of the differential cross section.  However, it is
a bit easier to just calculate $f_+(q^2=0)$ using lattice QCD and take
the experimental measurement~\cite{Moulson:2017ive} 
that determines $|V_{us}| f_+(0) = 0.21654(41)$.
The latest result for the form factor is~\cite{Bazavov:2018kjg}
\begin{equation}
f_+(0) = 0.9696(15)_{\rm stat}(11)_{\rm sys} = 0.9696(19)\;, 
\end{equation}
separating statistical and systematic errors before combining in quadrature.
The total theoretical error is 0.19\%,
the same size as the experimental
error.  Figure~\ref{fig:Form_factors} taken from \cite{Bazavov:2018kjg}
shows the result just quoted (denoted ``This work'') along with
FLAG 2016~\cite{Aoki:2016frl} averages in black and the results 
included in the averages
as green squares.  Blue circles come from non-lattice QCD calculations.
See \cite{Bazavov:2018kjg} for details.
Using the previously quoted experimental value~\cite{Moulson:2017ive}, we find
\begin{equation}
|V_{us}| = 0.22333(43)_{f_+(0)}(42)_{\rm exp} = 0.22333(60)\;.
\label{eq:vus}
\end{equation}

\begin{figure}[bth]
\begin{center}
\includegraphics[width=0.53\textwidth]{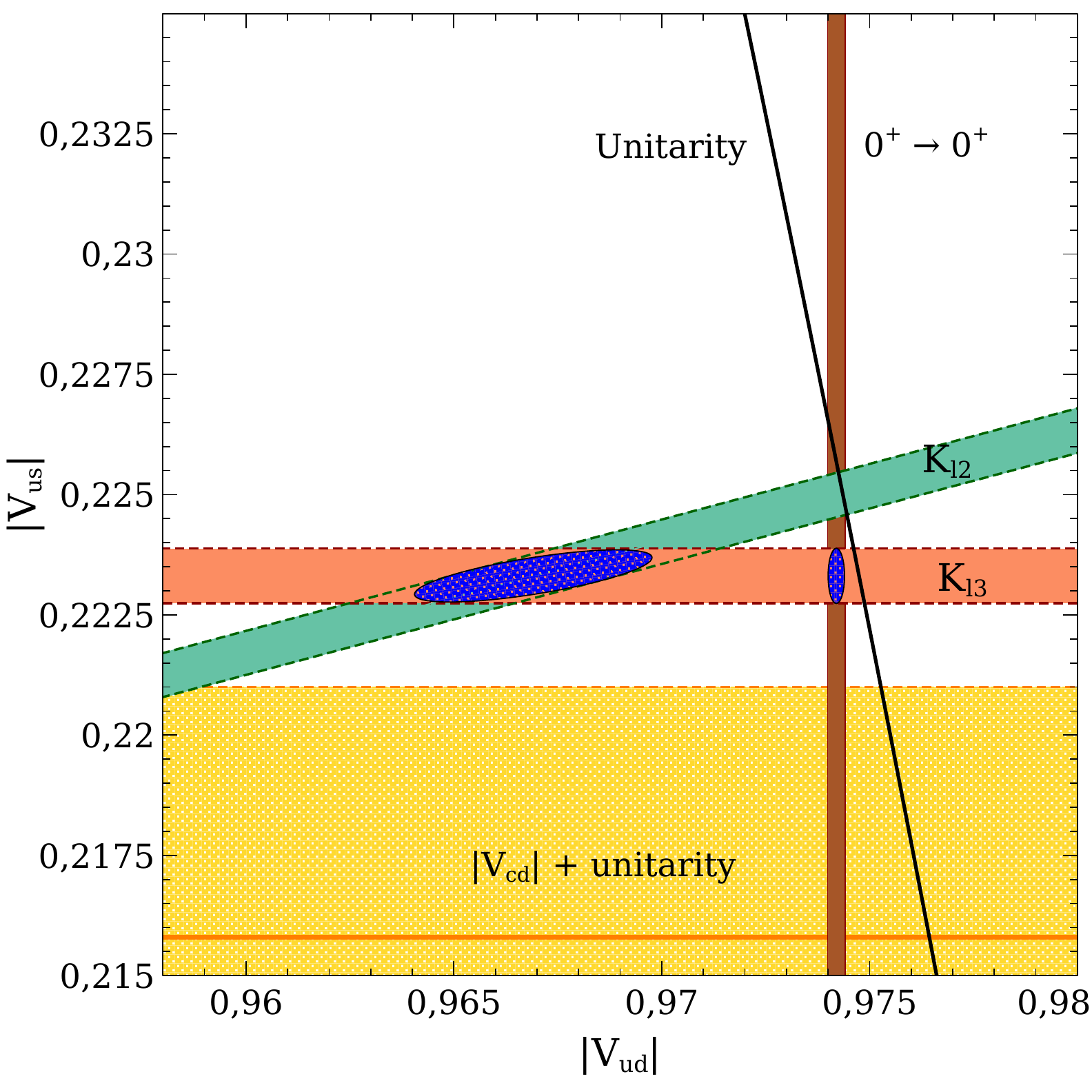}
\vspace{-5mm}
\caption{
\label{fig:Unitarity}
Constraints in the $|V_{ud}|$-$|V_{us}|$ plane from leptonic ($K_{l2}$) and
semileptonic  ($K_{l3}$) decays.  In addition, we have the result of
nuclear $\beta$-decay on $|V_{ud}|$, unitarity, and a wide horizontal
band from $|V_{cd}|$ combined with unitarity.  
(See \cite{Bazavov:2018kjg}.)
}
\end{center}
\end{figure}

We consider the implications of these results on first row unitarity
in Fig.~\ref{fig:Unitarity}.  Since $|V_{ub}|$ is so small, we can
neglect it.  The vertical band labeled $0^+\to 0^+$
comes from analysis of superallowed
nuclear $\beta$-decays~\cite{Hardy:2018zsb} and is independent of lattice
QCD input.  The diagonal band labeled $K_{l2}$ comes from the ratio of 
decay constants $f_K/f_\pi$, and the horizontal band 
labeled $K_{l3}$ comes from the kaon semileptonic form factor.  
The diagonal band, the vertical band,
and the unitarity curve nicely intersect.  However, there is tension
with unitary when we look at the kaon semileptonic decay.
The small blue ellipse uses the result in Eq.~\ref{eq:vus} and the value
of $|V_{ud}|$ from \cite{Hardy:2018zsb}, and we see that it does not intersect
the unitarity curve.  We find that
\begin{equation}
|V_{ud}|^2 + |V_{us}|^2 + |V_{ub}|^2 -1 =  -0.00104(27)_{V_{us}}(41)_{V_{ud}}
\end{equation}
which is $2.1\sigma$ from zero.
The large blue ellipse does not rely on $V_{ud}$ from $\beta$-decay and only
uses results from pion leptonic decay and kaon leptonic and semileptonic
decay.  There is a clear tension with unitarity.  In this case, we have
\begin{equation}
|V_{ud}|^2 + |V_{us}|^2 + |V_{ub}|^2 -1 =  -0.0151(38)_{f_+(0)}(35)_{f_K^\pm/f_\pi^\pm}(36)_{\rm exp}(27)_{\rm EM}
\end{equation}
which is $2.2\sigma$ different from zero.  The tension between leptonic
and semileptonic determination of $|V_{us}|$ and $|V_{ud}|$ can also be 
seen in the FLAG 2016~\cite{Aoki:2016frl} Fig.~7 summary plot 
for $|V_{us}|$ and $|V_{ud}|$
in which
semileptonic results are triangles and leptonic results are squares.  The
tension is most noticeable for the $N_f=2+1+1$ calculations where the
precision is higher.

\section{Second Row of CKM Matrix}
\begin{figure}[tb]
\begin{center}
\includegraphics[width=0.57\textwidth]{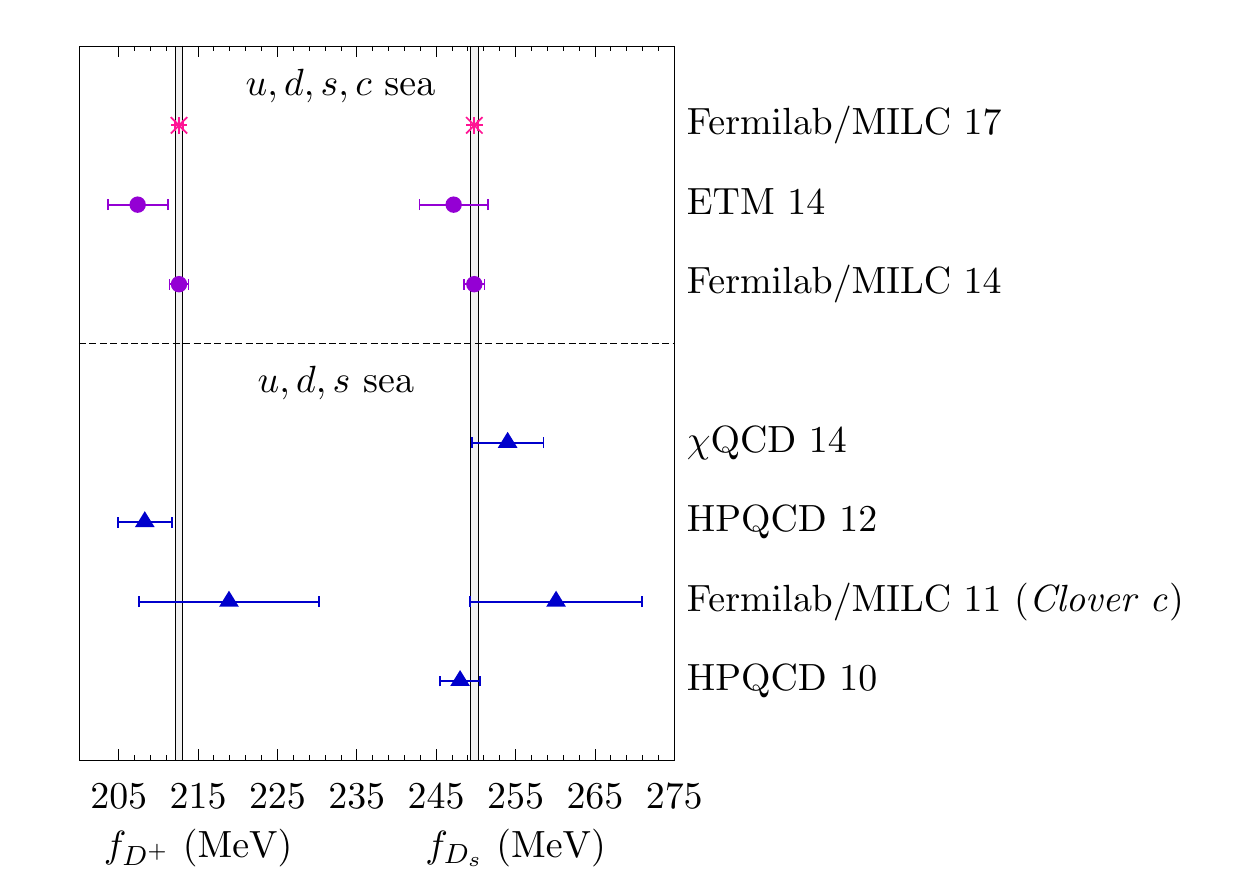}
\vspace{-5mm}
\caption{
\label{fig:fD_summary}
Comparison of recent calculations of $f_{D^+}$ and $f_{D_s}$ with $N+f=2+1+1$
and $2+1$.  From \cite{Bazavov:2017lyh}.
}
\end{center}
\end{figure}

We now turn to the decay of charm mesons to determine $|V_{cd}|$ and $|V_{cs}|$.
In 2005, the initial $N_f=2+1$ calculations of decay
constants were done with an accuracy of roughly 10\%.  The latest results for
$f_{D^+}$ and $f_{D_s}$ now have errors $<0.3$\%.  We have~\cite{Bazavov:2017lyh}
\begin{equation}
f_{D^+} = 212.7(0.6) \mathrm{MeV}, \qquad f_{D_s} = 249.9(0.4) \mathrm{MeV}\;.
\end{equation}
Prior to that calculation, the FLAG~\cite{Aoki:2016frl}
and Particle Data Group (PDG)~\cite{Patrignani:2016xqp} 
average values had errors of roughly
1--3\,MeV.  Figure~\ref{fig:fD_summary} summarizes the best recent 
calculations~\cite{Bazavov:2017lyh}.

We use experimental results from the PDG~\cite{Patrignani:2016xqp} to determine
the CKM matrix elements from the lattice QCD decay constants.
They have $f_{D}|V_{cd}|=45.91(1.05) \mathrm{MeV}$ and 
$f_{D_s}|V_{cs}|=250.9(4.0) \mathrm{MeV}$.  
The experimental error is 1.6--2.3\%.  For the CKM matrix element, we have
\begin{eqnarray}
|V_{cd}|_{{\rm SM},\, f_D}  &= 0.2152(5)_{f_D}(49)_{\rm expt}(6)_{\rm EM},\\
    |V_{cs}|_{{\rm SM},\, f_{D_s}} &= 1.001(2)_{f_{D_s}}(16)_{\rm expt}(3)_{\rm EM}\;,
\end{eqnarray}
where the errors are from lattice decay constant, experiment, and a structure
dependent electromagnetic correction.  The experimental errors are dominant.

Turning to the charm meson semileptonic decays, the FLAG 2016 form factor
average is based on HPQCD results from 2010~\cite{Na:2010uf} and 
2011~\cite{Na:2011mc}.  There are new results
from ETM Collaboration~\cite{Lubicz:2017syv} and JLQCD~\cite{Kaneko:2017xgg}.  
In addition, FNAL/MILC is completing an analysis that
should soon have errors smaller than those of HPQCD.  I have taken a rough
average of the three results above, even though they mix results with $N_f=2+1$
and $2+1+1$.  I find $f_+^{D\to\pi}(0) = 0.637(20)$ and 
$f_+^{D\to K}(0) = 0.745(15)$.  These values may be a little aggressive.
The FLAG 2016 values are 0.666(29) and 0.747(19), respectively.  Using
the HFLAV 2016 values~\cite{Amhis:2016xyh}  $f_+^{D\pi}|V_{cd}|= 0.1426(26)$ and $f_+^{DK}|V_{cs}|=
0.7226(34)$, we obtain $|V_{cd}|= 0.2239(76)$ and $|V_{cs}|=0.970(20)$
corresponding to errors of 3.4\% and 2.1\%.  In each case, the error is
dominated by the error in the lattice form factor input.  I updated the
experimental input after FPCP 2018, so the values here are different
from those in my slides.

We can test second row unitarity using a variety of determinations of 
$|V_{cd}|$ and $|V_{cs}|$.  In this case, $|V_{cb}| \approx 0.0414(8)$
contributes about 0.0017(6) to the unitarity sum.  We consider 
in Table~\ref{tab:secondrow} the latest result using leptonic decay constants from
\cite{Bazavov:2017lyh}, the FLAG 2016 $N_f=2+1+1$ result, the latest ETMC
semileptonic result~\cite{Riggio:2017zwh}, and my rough average of 
semileptonic results.  We find a slight ($1.5\sigma$) tension from
the leptonic decay determination and none from the semileptonic.  The
semileptonic error is dominated by experimental error and the semileptonic by
theory.  However, both will improve.
\begin{table}
\caption{Tests of second row unitarity from various determinations of
$|V_{cd}|$ and $|V_{cs}|$.
}
\label{tab:secondrow}
\begin{center}
\begin{tabular}{cl}
\hline
$|V_{cd}|^2 + |V_{cs}|^2 + |V_{cb}|^2 -1$ & input\\
$0.049(2)_{|V_{cd}|}(32)_{|V_{cs}|}(0)_{|V_{cb}|}$ & FNAL/MILC leptonic\cite{Bazavov:2017lyh}\\
0.06(3) & FLAG 2016 leptonic\\
-0.004(64) & ETMC semileptonic~\cite{Riggio:2017zwh}\\
0.005(53) & my semileptonic average\\[2pt]
\hline
\end{tabular}
\end{center}
\end{table}

\section{Decays of Hadrons with $b$ Quarks}
Decays of hadrons containing $b$ quarks have been studied in order to determine
$|V_{ub}|$ and $|V_{cb}|$.  Mesonic decays have been extensively studied by
a number of groups.  Recently, Meinel and his collaborators have been looking
at several decays of baryons with $b$ or $c$ quarks~\cite{Detmold:2015aaa}.  
Rare decays involving
flavor changing neutral currents (FCNC) are a good place to look for new
physics as FCNC processes vanish at the tree level.  These processes also
may involve third row CKM matrix elements and provide an alternative
to $B_{(s)}$ meson mixing for determining $|V_{td}|$ and $|V_{ts}|$.  Meson
mixing is covered by FLAG. 

\begin{figure}[tb]
\begin{center}
\includegraphics[width=0.58\textwidth]{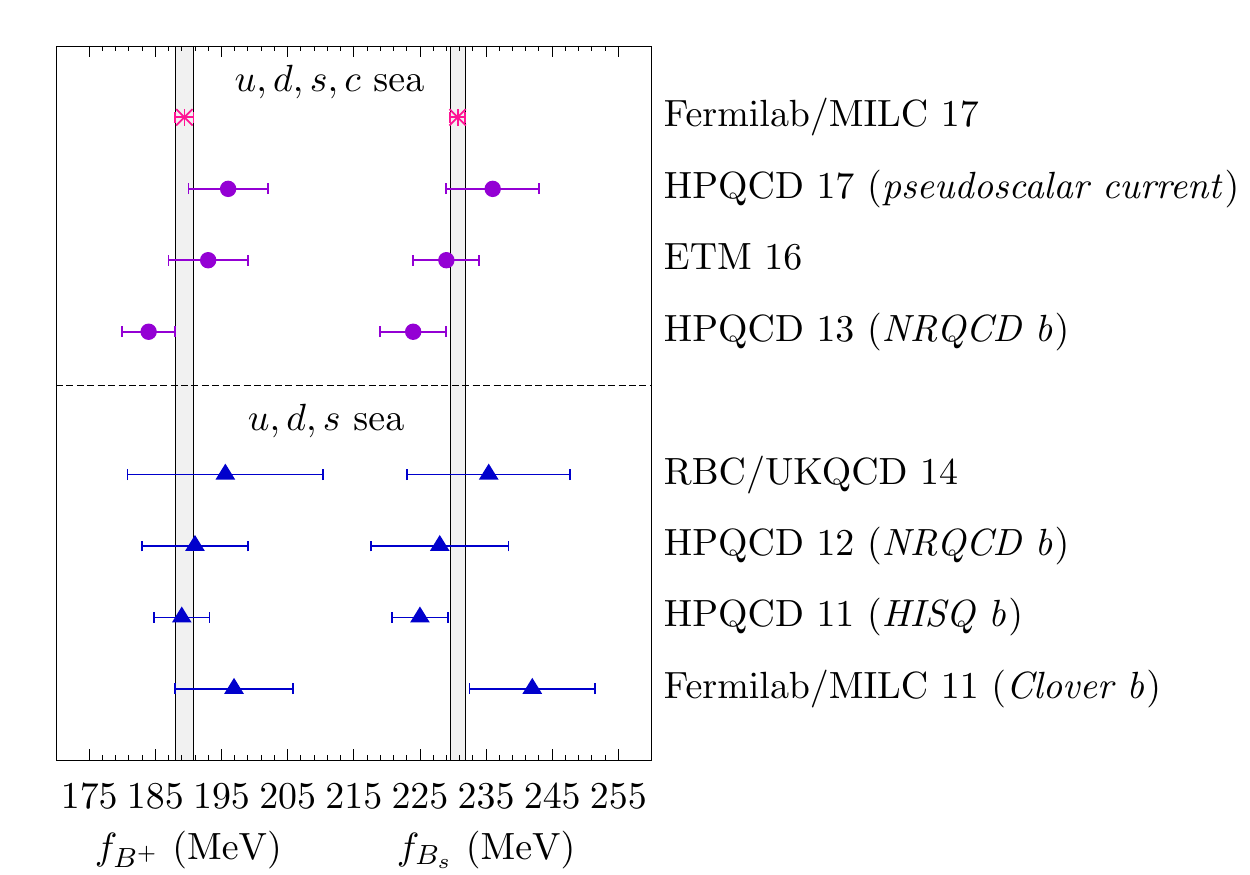}
\caption{
Comparison of recent calculations of $f_{B^+}$ and $f_{B_s}$ with $N_f=2+1+1$
and $2+1$.  From \cite{Bazavov:2017lyh}.
}
\label{fig:fB_summary}
\end{center}
\end{figure}

Reference~\cite{Bazavov:2017lyh} provides the best values for $B_{(s)}$
meson decay constants. (See Fig.~\ref{fig:fB_summary}).
Errors are $<1.3$\,MeV or 0.7\%.  There is good agreement
with earlier calculations that have errors as small as 5--7\,MeV.  To exploit
these results to get $|V_{ub}|$, we await precise results from 
Belle II for $B\to\tau\nu$, as the difference between BaBar and 
Belle is large~\cite{Patrignani:2016xqp}.

\begin{figure}[tb]
\begin{center}
\begin{minipage}{60mm}
\includegraphics[width=60mm]{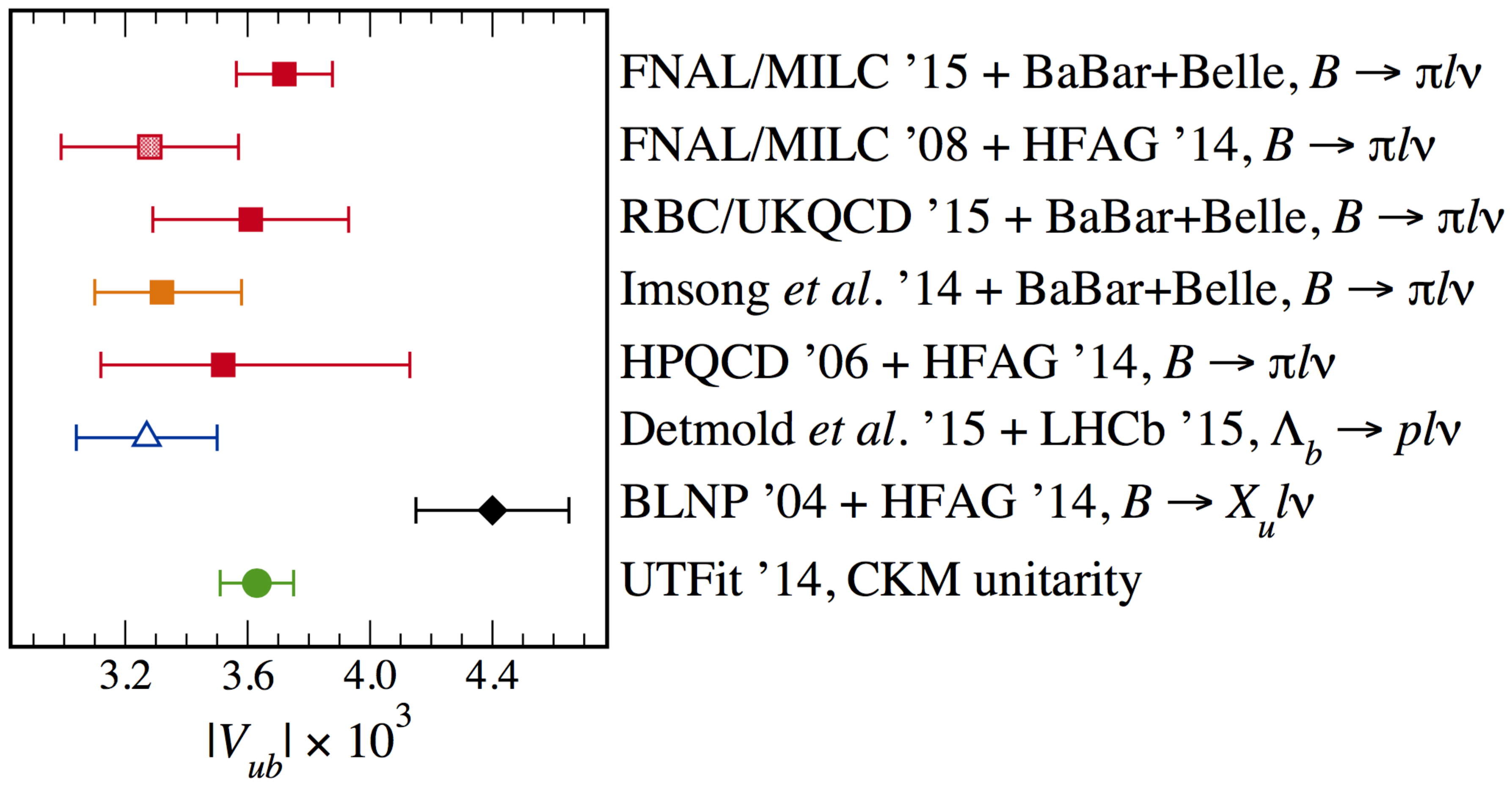}
\end{minipage}
\hspace{1mm}%
\begin{minipage}{60mm}
\includegraphics[width=60mm]{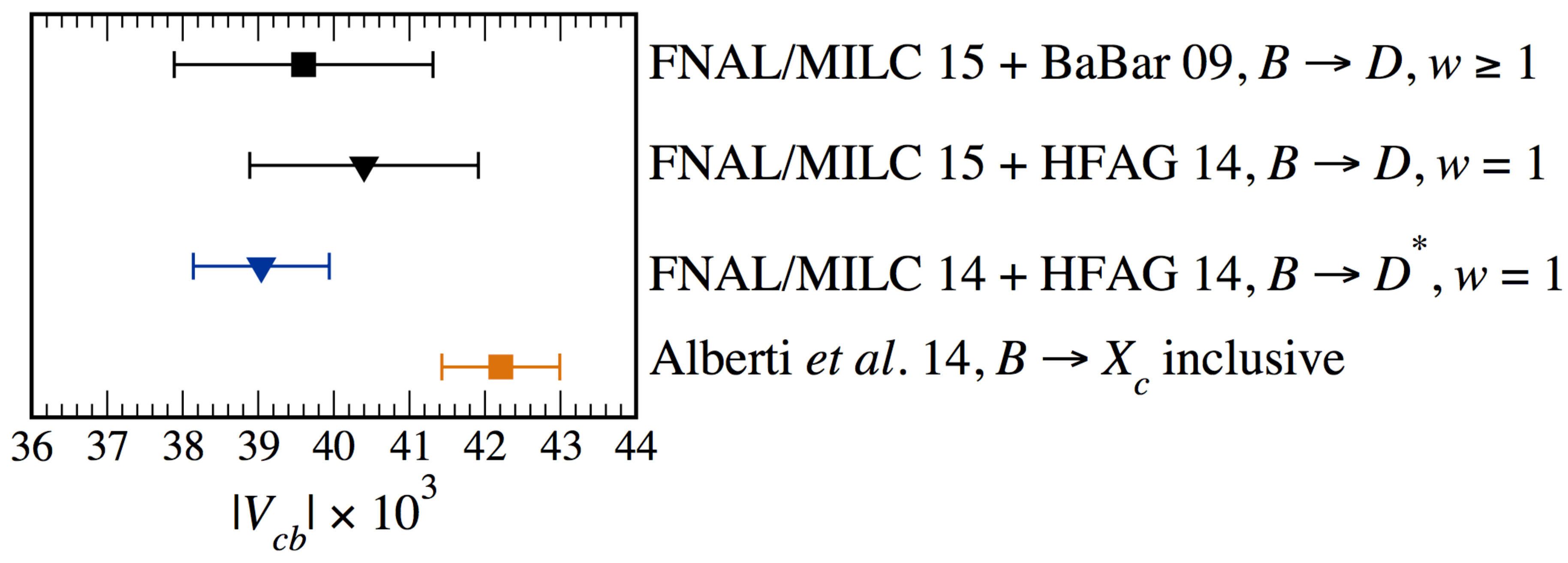}
\end{minipage}
\caption{
\label{fig:Vcb2015}
{\bf L} Comparison of several determinations of $|V_{ub}|$~\cite{Lattice:2015tia}.
{\bf R} Comparison of several determinations of $|V_{cb}|$~\cite{Lattice:2015rga}.
}
\end{center}
\end{figure}

Turning to bottom hadron semileptonic and rare decays, there are many possible
channels.  For $|V_{ub}|$ these include
$B\to\pi\ell\nu$, $B_s\to K\ell\nu$, $B_s\to K^*\ell\nu$, 
and $\Lambda_b\to p\ell\nu$.  For $|V_{cb}|$, we might study 
$B\to D\ell\nu$, $B\to D^*\ell\nu$, $B_s\to D_s^{(*)}\ell\nu$, and
$\Lambda_b\to \Lambda_c\ell\nu$.  We can test lepton universality as $\ell$
can be $e$, $\mu$, or $\tau$.  There are also interesting rare decays
such as $B^0\to \mu^+\mu^-$, $B_s\to \mu^+\mu^-$, $B_s\to\phi\ell\nu$, and $B\to K\ell^+\ell^-$.
Unfortunately, there is not enough time to cover all of these decays.  
Let's consider the long standing difference between 
CKM matrix elements as determined in exclusive and inclusive decay 
measurements.  
In 2015, the form factors needed for $B\to\pi\ell\nu$ were 
updated~\cite{Lattice:2015tia} and the resulting value of
$|V_{ub}| = 3.72(16)\times 10^{-3}$, was in somewhat better agreement
with the inclusive value as seen in Fig.~\ref{fig:Vcb2015}(L).
The figure shows several different determinations
of $|V_{ub}|$ including one based on $\Lambda_b$ decay (triangle)~\cite{Detmold:2015aaa}.  
The inclusive result is plotted as a diamond, unitarity as a circle.  
The exclusive value of $|V_{ub}|$ is in good agreement
with unitarity, but the inclusive one is not.  As mentioned
above, the leptonic decay $B\to\tau\nu$ could shed light on $|V_{ub}|$, but 
we'll need to wait for Belle II results for that.



As of 2015, the situation for $|V_{cb}|$ is depicted 
in Fig.~\ref{fig:Vcb2015}(R).
At that time, exclusive decay processes $B\to D^*\ell\nu$ and $B\to D\ell\nu$
were both being studied.  The experimental
error was larger in the $D$ channel (3.9\%) whereas it was just 
1.4\% for the $D^*$ channel.  In the figure, $w$ is an alternate kinematic
variable equivalent to $q^2$.  The form factor at $w=1$ (zero recoil)
can be calculated using lattice QCD; however, it is difficult to get the
corresponding experimental value as the differential decay
rate vanishes there, so it is necessary to fit the experimental results
as a function of $w$.  For the $D$ channel, the theoretical form factors
were available for a range of $w$.  The notation HFAG '14 in the figure
indicates that the experimental input for $w=1$ came from the fit of
the Heavy Flavor Averaging Group (now HFLAV).  
In 2016, Bigi and Gambino~\cite{Bigi:2016mdz}
used updated Belle data for $B\to D\ell\nu$ and the BGL parameterization~\cite{Boyd:1997kz}
to obtain $|V_{cb}| = (40.49\pm 0.97)\times 10^{-3}$.  In 2017,
Bigi, Gambino, Schacht~\cite{Bigi:2017njr}; 
and Grinstein and Kobach~\cite{Grinstein:2017nlq} examined new Belle 
data~\cite{Abdesselam:2017kjf} for $B\to D^*\ell\nu$ and found a 10\% difference when changing between
CLN~\cite{Caprini:1997mu} and BGL parameterizations of the experimental data.  Using CLN,
they found $(38.2\pm1.5)\times 10^{-3}$, and for BGL, they found
$(41.7\pm2.0)\times 10^{-3}$.  
The PDG inclusive value for $|V_{cb}| = (42.2\pm 0.8)\times 10^{-3}$.  We see that for $B\to D^*$, exclusive and
inclusive $|V_{cb}|$ values
are totally compatible, and for $B\to D$, the difference between
inclusive and exclusive determinations is only $1.36\sigma$.  Thus,
for $|V_{cb}|$, the issue is largely resolved (at least) until the errors can be
reduced.


%
%
\section{Conclusions and Outlook}
There has been very significant progress using lattice QCD to 
calculate hadronic matrix elements needed for precise evaluation
of SM contributions to numerious decay processes.
This theoretical input is essential to determine the CKM matrix.
A number of quantities can now be calculated to sub-percent
accuracy.  The interplay between theory and experiment will continue
to yield increasingly stringent tests of the SM.
In semileptonic decays, we see some tension
with unitarity in the first and second rows.  
In the first row,
we see slightly $>2\sigma$ tension with unitarity from semileptonic K decay.
There is some tension between leptonic and semileptonic determinations
of $|V_{ud}|$ and $|V_{us}|$.  The tests of unitarity in the second row
are not as stringent.  The difference between exclusive and inclusive
determination of $|V_{cb}|$ may be due to how the experimental data
had been fit; however, for $|V_{ub}|$ a difference remains.
Although I had hoped to cover
some of the other recent observations that are in tension with
the SM predictions, there was not enough time.  Future results
from Belle II, BES III, and LHCb, combined with increasingly
precise calculations from lattice QCD, will provide more critical
tests of the SM and opportunities to find evidence of new physics.

\paragraph{Acknowledgments}
I thank the FPCP organizers for their wonderful hospitality
and a stimulating conference.  I gratefully acknowledge my colleagues
in the Fermilab Lattice and MILC Collaborations for wonderful working
relationships and friendships.  I also thank
FLAG members who contribute countless hours to making lattice QCD results
more easily available to the wider community.  This work was supported by the
US DOE grant DE-SC0010120.

%
%

\end{document}